\documentclass[reprint,aps,prl,showpacs]{revtex4-1}
\usepackage{graphicx,epsfig}
\usepackage{epstopdf}
\usepackage{bm}

\newcommand{\3}{$^3$He}
\newcommand{\delR}{\delta \! R}
\newcommand{\delv}{\delta \! v}

\newcommand{\dif}{\, \mathrm{d}}

\begin{document}

\title{Reconnections of quantized vortex rings in superfluid $^4$He at very low temperatures}

\author{P. M. Walmsley$^a$}
\author{P. A. Tompsett$^a$}
\author{D. E. Zmeev$^{a,b}$}
\author{A. I. Golov$^a$}
\affiliation{$^a$ School of Physics and Astronomy, The University of Manchester, Manchester M13 9PL, UK}
\affiliation{$^b$ Department of Physics, Lancaster University, Lancaster, LA1 4YB, UK}
\date{\today}

\begin{abstract}
Collisions in a beam of unidirectional quantized vortex rings of nearly identical radii $R$ in superfluid $^4$He in the limit of zero temperature (0.05\,K) were studied using time-of-flight spectroscopy. Reconnections between two primary rings result in secondary vortex loops of both smaller and larger radii. Discrete steps in the distribution of flight times, due to the limits on the earliest possible arrival times of secondary loops created after either one or two consecutive reconnections, are observed. 
The density of primary rings was found to be capped at the value $500{\rm \,cm}^{-2} R^{-1}$ independent of the injected density. This is due to collisions between rings causing  piling-up of many other vortex rings.
Both observations are in quantitative agreement with our theory. 
\end{abstract}

\maketitle


Turbulence appears in various systems: fluids, plasmas, interstellar matter -- with common properties such as the existence of long-lived regions of concentrated vorticity, whose reconnections facilitate the evolution of the flow field and redistribution of the kinetic energy between length scales. 
A paradigm of an isolated vortical structure is a vortex ring \cite{ShariffLeonard1992, Donnelly}, and their pair interactions are a testbed of the physics of vortex reconnections. 
 Numerical simulations of collisions of two vortex rings predict various outcomes: either a single ring or several rings, depending on the initial conditions \cite{KoplikLevine96,Kivotides2003}. There were experimental attempts to visualize these processes in classical fluids \cite{Kambe1971, Fohl1975, Oshima1977}; however, they are often hard to interpret because of the inevitable decay, core instabilities and poor characterization of vortex rings in viscous fluids. 

Quantized vortex rings in superfluids have an advantage because they are slender and stable, and can  hence be well-characterized quantitatively \cite{RR1964}. Recently, there were many theoretical investigations into reconnections of quantized vortex lines  \cite{Kerr87, KoplikLevine93, KoplikLevine96, Nore97, Kivotides2001, Leadbeater2001, Leadbeater2003, Chatelain2003, Kivotides2003, Mitani2006, Nemirovskii2006, Barenghi2006, Alamari2008, Kerr2011, Kursa2011, Hanninen2013, Caplan2014}. In particular, for acute angles between two antiparallel reconnecting vortex lines the generation of a cascade of small vortex rings was predicted \cite{Svistunov1995,Kerr2011,Kursa2011}. The outcome is reminiscent of that of the Crow instability \cite{Crow} of antiparallel vortices observed in air. Experimentally, reconnections of vortex lines in superfluid $^4$He have been visualized \cite{Beweley2008,Fonda2014} but only at high temperatures when vortex motion is damped. 
 Reconnections of vortex loops comprising a vortex tangle, i.\,e. {\it Quantum Turbulence} (QT) \cite{Schwarz1988, VinenNiemela2002}, especially those leading to the emission of vortex rings \cite{Yano2013}, are an important mechanism of redistributing energy towards smaller length scales in QT  \cite{Svistunov1995, Tsubota2000, BarenghiRingEmission2002, NazarenkoReconnection, Nemirovskii2008, Nemirovskii2010, KS, Yamamoto2012, Kondaurova2012,  Baggaley2013, Salman2013}.  

 In superfluid $^3$He-B \cite{Bradley2005} and $^4$He \cite{PRLWalmsley2008}, collisions and subsequent reconnections, in a dense beam of vortex rings generate QT. 
Longer and more intensive beams of rings result in tangles that show large-scale velocity fluctuations \cite{TsepelinFluctuations, Roche} and the late-time decay 
\cite{Bradley2005,PRLWalmsley2008,WalmsleyLongInjection,WalmsleyPNAS2014}, both characteristic of classical turbulence. This implies the existence of the {\it inverse cascade} of energy from the small length scales (of order ring radii) into which the initial energy is injected -- up to the size of the resulting tangle. It was speculated \cite{Baggaley2013} that the inverse cascade might be maintained by the merger of pairs of rings into  larger loops. Yet, no direct quantitative observations of ring-ring reconnections have been reported so far.

 This Letter reports the first quantitative observations of either one or two consecutive reconnections, and the discovery of the ensuing universal state of depleted density -- within a beam of unidirectional quantized vortex rings all of similar radii, with their number density $n$ and radius $R$ under our control. The resulting mechanism of seeding the large-scale velocity fluctuations out of a seemingly random beam of vortex rings is suggested.
 
In our experiments, to create vortex rings of a required size and to detect their arrival, each was tagged by an excess electron trapped on the vortex core. Applying an electric field  along the $x$-axis allowed small seed charged vortex rings (CVRs), injected at $x=0$, to grow to the desired radius $R(x)$, and also to trace the location of the reconnection process that resulted in small secondary charged vortex rings. 
The radius of a quantized vortex ring is directly related to its self-induced velocity, $v\sim \kappa/R$ -- which determines its arrival time at the collector at $x=d$.  The numbers of primary and secondary vortex rings as a function of their radii could be extracted from the time-dependence of the collector current $I_{\rm c}(t)$ through time of flight  spectroscopy. 
 The radius of primary rings at the collector, $R(d)$, was varied within 1--6\,$\mu$m, with number density $n(d)$ between $10^{4}$ and $10^{7}$\,cm$^{-3}$; while the mean radius of the seed CVRs was estimated as $\bar{R_0} \leq 0.5$\,$\mu$m.

 The energy of a CVR, subject to a potential $\phi(x)$, is  
  ${\cal E}(x) = {\cal E}_0 + e \phi(x)$ in the absence of dissipation at $T<0.5$\,K. 
The velocity and energy depend on $R$ \cite{Donnelly},
\begin{equation}
 v = \frac{\kappa}{4\pi R}\left(\Lambda(R) - \frac{1}{2}\right), {\rm \,\,}
 {\cal E} = \frac{\kappa^2 \rho R}{2}\left(\Lambda(R) - \frac{3}{2}\right),
\label{vE}
\end{equation}
 where $\kappa = h/m_4$ is the circulation quantum, $\rho$ is the density of superfluid, $\Lambda = \ln\frac{8R}{a_0}$ and $a_0 = 1.3$\,\AA \cite{GlabersonDonnelly1986}. 
 

A deeper insight can be gained within the approximation for constant $\Lambda \approx 13$ and uniform field $\phi(x) = \frac{U}{d}x$. The radius of a  CVR then grows linearly with $x$, \begin{equation}
R(x) \approx R_0 + \frac{2eU}{\rho\kappa^2(\Lambda-3/2)}\frac{x}{d},
\label{R_appr}
\end{equation}
and the time for a CVR to travel from $x=0$ to $x=d$,
\begin{equation}
\tau_1 \approx \frac{4\pi d}{\kappa (\Lambda -1/2)}R_0 + \frac{4\pi e d}{\rho\kappa^3 (\Lambda - 1)^2}U,
\label{tau1}
\end{equation} 
increases with energy $eU$ because CVRs slow down as they expand (see Eq.\,\ref{vE}). In what follows, unless specified, we will be using the approximation $R_0 = 0$.

With increasing density of CVRs, collisions become more frequent. These collisions are caused by small fluctuations in the direction and magnitude of the rings' velocities -- mainly due to the variations in initial radii $\delta R_0$ and direction of the seed CVRs when injected at $x=0$. Along with reconnections upon a direct collision, hydrodynamic dipole-dipole interactions between neighboring CVRs (that grows in strength with increasing $n$ and $R$ -- and are hence the strongest near the collector at $x \rightarrow d$) affect the CVR's velocities.  The Coulomb repulsion between neighboring CVRs of $R>$\,1\,$\mu$m is much weaker than their hydrodynamic interaction.

A reconnection of two CVRs results in secondary vortex loops, which are generally non-circular. 
The two trapped electrons are now carried by either one or by two (if any) of the secondary rings. 
One special case allows an exact analysis of the consecutive trajectory of one of the electrons -- when a secondary CVR has a small initial radius ($R_0 \ll R$). Then its initial deformation and direction of motion can be disregarded, because, under the pull of the electric field, it quickly gains sufficient energy and impulse along the $x$-direction, so to a good accuracy can be treated as a circular vortex ring \cite{SamuelsDonnelly1991,CVRTsubota}. If such singly-charged loops are created after collisions at some $x=x_1$, their arrival at the collector ($x=d$) at time $\tau_2(x_1) = \tau_1\left[\left(\frac{x_1}{d}\right)^2+\left(1-\frac{x_1}{d}\right)^2\right]$ will be earlier than of any other CVRs with either larger initial size (slower) or double charge (more energetic, hence, slower). 
The earliest arrival time,
\begin{equation}
\tau_2^* \equiv \min(\tau_2)= \frac{\tau_1}{2} ,
\label{tau2}
\end{equation}
 will be for collisions at $x_1=d/2$. 
Furthermore, if a secondary small ring grows and then reconnects with another vortex loop at some point $x=x_2$ ($x_1<x_2<d$), and this creates a new small singly-charged ring, the latter will arrive at the collector at time $\tau_3(x_1,x_2) = \tau_1\left[\left(\frac{x_1}{d}\right)^2 +\left(\frac{x_2}{d}-\frac{x_1}{d}\right)^2  +\left(1-\frac{x_2}{d}\right)^2\right]$. The earliest arrival, at time 
\begin{equation}
\tau_3^* \equiv \min(\tau_3)= \frac{\tau_1}{3} ,
\label{tau3}
\end{equation}
 of these second-generation secondary CVRs will correspond to two reconnections at $x_1=d/3$ and $x_2=2d/3$.

The experimental cell \cite{WalmsleyJLTPPobell}, a cube-shaped volume of side $d=4.5$\,cm, was filled with isotopically-pure liquid $^4$He \cite{pure4He} at pressure 0.1\,bar and temperature 0.05\,K (see inset in Fig.\,1). 
Seed CVRs were injected through a gridded opening in the center of one plate. They then traveled along the axis of the container ($x$-axis) towards the center of the opposite plate to the collector electrode, placed behind a Frisch grid of radius $r = 6.5$\,mm and geometric transparency $\theta=0.92$. 
All currents and potentials are quoted with the opposite sign as if electrons had {\it positive} charge $e$. CVRs were subject to the propelling field set by potentials of plates $\phi(0)=0$ and $\phi(d)=U$, thus gaining energy $eU$ while travelling between the injector and collector grids \cite{CommentBuoyantRings}. The dependence $\phi(x)$ was close to the linear $\phi = U\frac{x}{d}$ (see inset in Fig.\,2). 

The seed CVRs resulted from reconnections within the dense vortex tangle, generated by the current of electrons emitted from a  tungsten tip \cite{GolovIshimoto} behind the injector grid through the voltage $U_{\rm tip}$. 
 These seed CVRs are injected in a broad range of angles; however, the impulse gained from the strong driving field quickly forces them to travel in nearly the same $x$-direction with a relatively narrow distribution of radii. 
The intensity and duration of the injected pulse were controlled by adjusting $U_{\rm tip}$ and its duration $\Delta t$ (all data presented here are for $\Delta t = 0.2$\,s). 
 For the same $U_{\rm tip}$ and $\Delta t$, the total charge injected through the grid was increasing with increasing drive voltage $U$ nearly linearly for all studied voltages. To quantify the time of flight $\tau_1$ of CVRs, we take the time interval between the middle of the tip voltage pulse and the position of the maximum of $I_{\rm c}(t)$ (and subtract the electronics response time  of 0.03\,s). 
 

 \begin{figure}
\includegraphics[width=7cm]{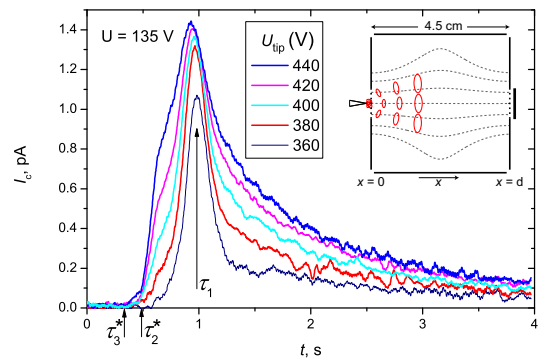}
\caption{(color online) Records of collector current, all for the same drive voltage $U=135$\,V, but different tip voltages $U_{\rm tip}$. The theoretical arrival time for primary CVRs, $\tau_1$ (Eq.\,\ref{tau1}), and of the earliest arriving secondary CVRs of first generation, $\tau_2^*$ (Eq.\,\ref{tau2}), and of second generation, $\tau_3^*$ (Eq.\,\ref{tau3}), are shown by arrows. Inset: Experimental cell with electric field pattern.}
\label{FigIvsTvarUtip2}
\end{figure}

Typical records of the collector current, $I_{\rm c}(t)$, following the injection of a pulse of CVRs are shown in Fig.\,\ref{FigIvsTvarUtip2}. These are all for the same drive voltage $U=135$\,V but several different injection currents. There is a well-defined peak at time $\tau_1 \approx 1.0$\,s corresponding to the arrival of primary CVRs. With increasing  density of CVRs, this peak initially grows in magnitude while maintaining its position, $\tau_1$, and width, $\sim \Delta t$. At higher numbers of injected CVRs, however, a broad pedestal begins to grow, coexisting with the original peak (whose magnitude is now saturated). This broad peak-pedestal is due to the secondary CVRs that result from collisions between primary CVRs. 
The current at $t > \tau_1$ reflects the arrivals of larger secondary vortex loops, while 
that at the earlier arrival times $t < \tau_1$ are from smaller secondary CVRs. 
A sharp step builds up at $\tau_2^*=\tau_1/2$, coinciding with the earliest possible arrival of the first generation of secondary CVRs (Eq.\,\ref{tau2}). At the highest intensity of injection, another sharp step begins to form at $\tau_3^*=\tau_1/3$, corresponding to the earliest possible arrival of the second generation of secondary CVRs (Eq.\,\ref{tau3}).

 \begin{figure}
\includegraphics[width=7cm]{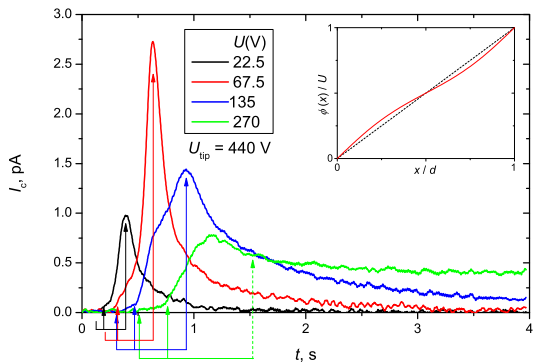}
\caption{
(color online) Records of collector current, all for the same $U_{\rm tip}=440$\,V, but different values of $U$. The right, middle and left arrows of the corresponding color point at the arrival time of primary CVRs, $\tau_1$, and the theoretical earliest arrival times for secondary CVRs of first (Eq.\,\ref{tau2}) and second  (Eq.\,\ref{tau3}) generation, respectively. 
The values of $\tau_1$ are defined as the positions of sharp peaks, except for $U=270$\,V -- where the peak due to primary CVRs is swamped by the broad pedestal due to secondary CVRs, hence the theoretical value of $\tau_1$ (see Fig.\,3) is used. For $U=22.5$\,V and 67.5\,V, the signal is too faint for the step at $\tau_3^*$ to be observable. Inset: The electrostatic potential $\phi(x)$ along the cell's axis.}
\label{FigIvsTvarU}
\end{figure}

In Fig.\,\ref{FigIvsTvarU}, we show $I_{\rm c}(t)$, similar to those in Fig.\,\ref{FigIvsTvarUtip2}, but now for the same tip voltage $U_{\rm tip} = 440$\,V and four different drive voltages $U$. With increasing $U$, the position of the peak $\tau_1$ increases as expected for isolated CVRs, Eq.\,\ref{tau1}. The peak's magnitude $I_{\rm m}(U)$ initially grows with $U$ but then, above $U=68$\,V, decreases -- even though the total collected charge $Q_{\rm c} = \int_0^\infty I_{\rm c}(t) \dif t$ keeps increasing. Simultaneously, the broad pedestal due to secondary CVRs progressively overgrows the primary peak until completely swamping it at $U=270$\,V. The sharp steps due to the earliest possible arrivals of secondary CVRs of first generation at $\tau_1/2$ (Eq.\,\ref{tau2}) and second generation at $\tau_1/3$ (Eq.\,\ref{tau3}) are labeled by arrows. 
We thus obtained quantitative evidences of either single or two consecutive reconnections of CVRs during their motion from the injector to collector. Furthermore, the substantial contribution to the collector current right after the cut-offs indicates that very small CVRs are created with high probability. This might contradict the expectations that reconnections result in vortex loops of size {\it comparable} to the radius of curvature of the initial vortex lines \cite{BarenghiRingEmission2002,KS}, but would support the picture of  a cascade of small vortex rings created by large-amplitude Kelvin waves generated after a reconnection of nearly antiparallel vortex lines when dissipation is small \cite{Svistunov1995,Kerr2011,Kursa2011}. 

 \begin{figure}
\includegraphics[width=7cm]{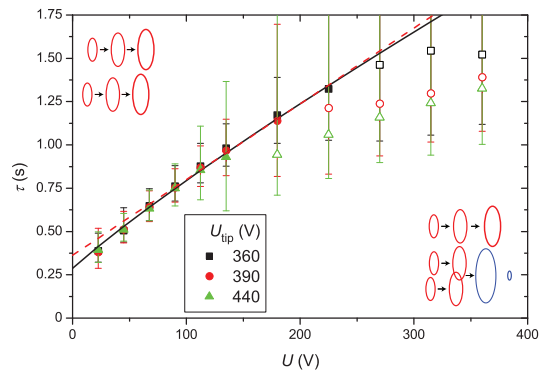}
\caption{(color online) The elapsed times between the middle of the emitter pulse and maximum of $I_{\rm c} (t)$ vs. $U$.  Closed symbols correspond to resolvable sharp peaks due to primary CVRs; open symbols are for the broadened peaks due to secondary CVRs where sharp peaks due to primary CVRs are no longer visible. Vertical bars indicate the peak width at half-maximum. The solid line shows the theoretical arrival times $\tau_1$ (Eqs.\,1--2) for  $R_0=0.5$\,$\mu$m. The dashed line shows the approximate solution for $\Lambda = 13$ and effective $R_0=0.8$\,$\mu$m (Eq.\,\ref{tau1}). Cartoons illustrate the expansion and progression of: (left) isolated primary (red) CVRs at low density, and (right) a reconnection resulting in secondary (blue) CVRs.} 
\label{FigTvsU}
\end{figure}

In Fig.\,\ref{FigTvsU}, the experimental arrival times $\tau_1(U)$ for several intensities of injection are plotted. For small drive voltages $U$ (i.\,e. when the radii of CVRs $R(d)$ and density of CVRs $n(d)$ are small) the experimental points agree with the theory for isolated CVRs (from Eqs.\,\ref{vE}--\ref{R_appr}). To characterize the range of the distribution of times of flight, the vertical bars show the width of $I_{\rm c}(t)$ at the $0.5 I_{\rm m}$ level. One can see that at small $U$ the width is constant, being equal to the injection duration $\Delta t = 0.2$\,s. Above a certain value of $U \sim 100$\,V (that decreases with increasing injection intensity, $U_{\rm tip}$), the collector pulse broadens and the position of the maximum of $I_{\rm c}(t)$ no longer agrees with the theoretical prediction for isolated CVRs (this coincides with the complete disappearance of the sharper peak due to primary CVRs, as on the trace for $U=270$\,V in Fig.\,\ref{FigIvsTvarU}); secondary CVRs dominate $I_{\rm c}(t)$ in these conditions of high $n$ and $R$.

 \begin{figure}
\includegraphics[width=7cm]{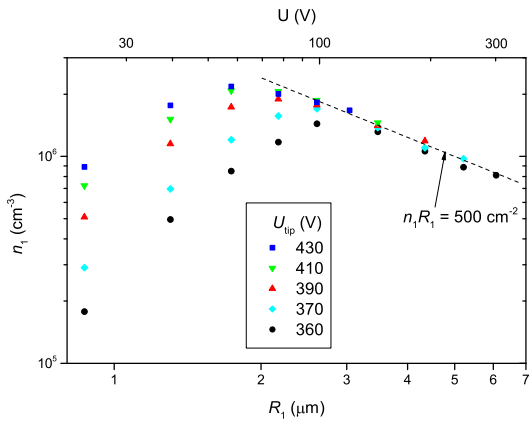}
\caption{(color online) Density of primary CVRs $n$ vs. their radius $R$ at $x=d$ measured for several different drive voltages $U$ (top axis) and injector tip voltages, $U_{\rm tip}$ (see legend).
}
\label{FigNvsR}
\end{figure}

The density of primary rings reaching the collector, $n_1 \equiv n(d)$, can be found from  the value of the collector current at its maximum (but only for the records $I_{\rm c}(t)$ that have a sharp peak at $t=\tau_1$, dominant over the pedestal due to the secondary CVRs),  
\begin{equation}
  I_{\rm m} \approx \theta \pi r^2 e  n_1 v_1 \approx \frac{1}{8}\theta r^2\rho\kappa^3(\Lambda-1)^2\frac{n_1}{U},
\label{Im}
\end{equation} 
where the relation $v_1 \equiv v(d) \approx \frac{\rho\kappa^3(\Lambda-1)^2}{8\pi eU}$ (from Eqs.\,\ref{vE}--\ref{R_appr}) has been used. 
In Fig.\,\ref{FigNvsR}, we plot $n_1$ (calculated from the experimental values of $I_{\rm m}(U)$ using Eq.\,\ref{Im}) vs. $R_1 \equiv R(d)$, the radii of primary CVRs near the collector, (calculated from Eq.\,\ref{R_appr} with $R_0=0$). 
Again, there are two regimes: at low $R_1$, $n_1$ increases with an increase of either $U_{\rm tip}$ or $U$, but at large $R_1$, $n_1$ becomes a {\it decreasing} function of $U$, independent of $U_{\rm tip}$. Yet, the total injected charge (as measured by the integral of the collector current) increases monotonically at all $U$. 
I.\,e., only at small $R_1$ and $n_1$, the CVR's density can be controlled by varying the injection current. At high $R_1$, only a small fraction of charge arrives with primary CVRs of now {\it universal} density $n_1 \approx 500{\rm \,cm}^{-2} R_1^{-1}$;  the rest (secondary CVRs) contribute to the broad pedestal in $I_{\rm c}(t)$ (Fig.\,\ref{FigIvsTvarUtip2}--\ref{FigIvsTvarU}).


The density of primary rings in the beam $n(x,t)$, along the trajectory $x(t)$, evolves according to Eq.\,11 in \cite{Supplement}: 
\begin{equation}
\frac{\dif n}{\dif t} = - n\frac{\dif v}{\dif x}  - f,
\label{RateEquation}
\end{equation}
where $-f$ is the  rate of losses, per unit volume and time, due to ring-ring collisions. For small injected density and radii, the collisions can be neglected, $f=0$, and the solution for the density near collector is  
\begin{equation}
n_1 =  \frac{8\pi I_m U}{\theta\pi r^2  \rho \kappa^3 (\Lambda -1)^2},
\label{n1}
\end{equation}
i.\,e. it can be varied by changing either the injected density of CVRs (characterized by $I_m$) or drive voltage $U$ -- as observed in the experiment. When collisions become rife at higher densities and radii, accounting for the removal of primary CVRs due to their binary collisions results in 
\begin{equation}
n_1R_1 = \frac{3}{\sigma_1 ' \delta \! R d} = 4\times 10^3{\rm\,cm}^{-2}, 
\label{nR1}
\end{equation}
where we used $\delta \! R = 0.5$\,$\mu$m and the geometric cross-section for collisions $\sigma_1 = \sigma_1 ' R^2$ with $\sigma_1 ' = 4\pi$. Furthermore,  the subsequent removal of  primary CVRs that bump into the slower loops \cite{CommentPile-ups} left after the described collisions of primary CVRs will result in the solution,
\begin{equation}
n_1R_1=\left( \frac{40\pi}{(\Lambda-1/2)\kappa \sigma_1 ' \sigma_2 ' \delta \! R \Delta t d} \right)^{1/2} = 1\times 10^3{\rm\,cm}^{-2},
\label{nR2}
\end{equation}
where the geometric cross-section for these pile-ups is estimated as $\sigma_2 = \sigma_2 ' R^2$ with $\sigma_2 ' = 4\sqrt{2}\pi$. 
Both  Eq.\,\ref{nR1} and Eq.\,\ref{nR2}  reproduce the experimental universal dependence $n_1R_1 = 500$\,cm$^{-2}$ qualitatively, while Eq.\,\ref{nR2} is actually quite close {\it quantitatively} (our 1-dimensional model underestimates $f$ by disregarding the transverse component of the relative motion of primary CVRs and hence overestimates the value of $n_1R_1$ by a factor of 2 or so \cite{Supplement}).

The fact of occasional piling-up of many primary CVRs, in turn, helps to explain the appearance of large-scale velocity fluctuations in the ensuing vortex tangle.
 In the initial random beam of primary CVRs, the fluctuations of the coarse-grained velocity on the length scales greater than $\sim n^{-1/3}$ (``quasi-classical'' flow) are small; the energy spectrum is concentrated around the small length scale of order $R$. 
The small secondary vortex rings, observed in this work, and Kelvin waves excited by reconnections are evidence of the direct cascade of energy towards smaller length scales \cite{Svistunov1995, Tsubota2000, BarenghiRingEmission2002, NazarenkoReconnection, KS, Nemirovskii2010}. 
 However, following any of the pile-ups of many vortex rings, strong fluctuations of the coarse-grained velocity field on the ``quasi-classical'' length scales $\gg n^{-1/3}$ are being  created. 
 This is the inverse cascade of energy in this strongly anisotropic system  \cite{Yamamoto2012, Baggaley2013}.


To summarize, we obtained quantitative evidence for collisions and reconnections of pairs of unidirectional vortex rings of similar radii that result in the creation of vortex loops of unequal size, including many small ones. We observed discrete steps at the time dependence of the collector current, that correspond to the earliest arrivals of the first and second generations of secondary CVRs. As each collision can cause a  removal of many primary vortex rings, increasing the density of injected CVRs results in a new  state in which the density of primary vortex rings is maintained at the critically depleted level independent of their initial density. The larger loops produced in the collisions become the seeds of quasi-classical QT with large-scale flow structures, which appear out of a seemingly random  beam of small quantized vortex rings.
 

This work was supported by the Engineering and Physical Sciences Research Council (Grants No. GR/R94855, EP/H04762X/1 and EP/I003738/1). 

\section{Supplementary material: density of unidirectional vortex rings subject to reconnections}

\subsection{Introduction}

In these supplementary notes, we describe effects of collisions of unidirectional charged vortex rings (CVRs) due to the small variations, $\sim \delv$, of their velocities around the mean value $v$. Any such collision of primary CVRs generally results in secondary CVRs of quite different size -- hence depleting the number of the primary CVRs. 
We consider purely one-dimensional trajectories of CVRs, $x(t)$, between the injector at $x=0$ and collector at $x=d$ \cite{1-d}. We assume a uniform electric field $U/d$, a negligible initial radius of CVRs, $R_0 \rightarrow 0$, and a constant logarithmic parameter $\Lambda \equiv \ln \frac{8R}{a_0} \approx 13$ (where $a_0 \approx 1.3$\,\AA\ is the core radius). Within the time of flight from the injector to collector, $\tau_1$, the lengthening of the beam, caused by the range of velocities $\delta v$, does not exceed the duration of injection $\Delta t = 0.2$\,s and is, hence, neglected.

The velocity and energy of a quantized vortex ring of radius $R$ carrying one electron of charge $e$:
\begin{equation}v = \frac{\kappa}{4\pi R}\left(\Lambda-\frac{1}{2}\right),\label{v(R)}\end{equation}
\begin{equation}{\cal E} = \frac{\kappa^2 \rho R}{2}\left(\Lambda-\frac{3}{2}\right) = e\frac{U}{d}x,\label{E(R)}\end{equation}
where $\rho=0.145$\,g\,cm$^{-3}$ is the density of superfluid helium and $\kappa = 1.0\times 10^{-3}$\,cm$^2$\,s$^{-1}$ is the quantum of circulation. 


The radius grows linearly with  $x$,
\begin{equation}R =  R_1\frac{x}{d},\label{R(x)}\end{equation}
where the radius of CVRs at collector, $x=d$, is 
\begin{equation}R_1 = \frac{2eU}{\rho\kappa^2(\Lambda-3/2)}\label{R1}.\end{equation}
 The time of flight between $x=0$ and $x=x_1$ is
\begin{equation}t(x) = \int_0^{x} v^{-1}dx' = \tau_1\frac{x^2}{d^2},\label{t(x)}\end{equation} 
i.\,e. the trajectory of primary CVRs is 
\begin{equation}x(t) = d\left(\frac{t}{\tau_1}\right)^{1/2},\label{x(t)}\end{equation}
where  the time of flight from injector to collector is 
\begin{equation}\tau_1 = \frac{4\pi e d}{\rho\kappa^3(\Lambda-1)^2}U = \frac{2\pi d}{\kappa(\Lambda-1/2)}R_1.\label{tau1}\end{equation}


 The velocity field, its gradient and pulse length in $x$-space can be conveniently expressed (including the explicit time dependence along the trajectory $x(t)$):
\begin{equation}v\equiv \frac{\dif x}{\dif t} =\frac{d^2}{2\tau_1}x^{-1} = \frac{d}{2}\tau_1^{-1/2}t^{-1/2},\label{v(x)}\end{equation}
\begin{equation}\nabla v = -\frac{d^2}{2\tau_1} x^{-2} = -\frac{1}{2}t^{-1},\label{nablav}\end{equation}
\begin{equation}\Delta x = v\Delta t = \frac{d^2\Delta t}{2\tau_1}x^{-1} = \frac{d\Delta t}{2}\tau_1^{-1/2}t^{-1/2}.\label{Deltax}\end{equation}


\subsection{Evolution of the density of CVRs}

The number density of primary rings in the beam $n(x,t)$, along the trajectory $x(t)$, evolves according to: 
\begin{equation}
\frac{\dif n}{\dif t} \equiv \frac{\partial n}{\partial t} + v \frac{\partial n}{\partial x} = \left(-\frac{\partial (nv)}{\partial x} -f \right) + v \frac{\partial n}{\partial x}   = - n \frac{\dif v}{\dif x}  - f,
\label{RateEquation}
\end{equation}
where $-f$ is the rate of losses due to ring-ring collisions. 

For small $n$, at which $f\rightarrow 0$, the number of CVRs is conserved, and the solution is 
 $nv = const$, i.\,e. $n = A x$. 
Experimentally, the value of constant $A$ can be determined from the magnitude of the peak of the collector current due to primary CVRs, 
\begin{equation}I_m = \theta\pi r^2 e  n_1 v_1 = \theta\pi r^2 e  A d^2 (2\tau_1)^{-1},\label{Im}\end{equation} 
where $r$ and $\theta$ are the radius and transparency of the collector grid, $n_1\equiv n(d)$ and $v_1\equiv v(d)$. 
We arrive at 
\begin{equation}
n(x) = \frac{2\tau_1 I_m}{\theta\pi r^2 e  d^2} x = \frac{8\pi I_m U}{\theta\pi r^2  \rho \kappa^3 (\Lambda -1)^2d} x.
\label{n1}
\end{equation}
The density at collector is 
\begin{equation}
n_1 =  \frac{8\pi I_m U}{\theta\pi r^2  \rho \kappa^3 (\Lambda -1)^2}.
\label{n1}
\end{equation}
Thus, in this regime of conserved primary CVRs, their density can be varied by changing either the intensity of injection (as measured by $I_m$) or drive voltage $U$. 

\subsection{Collisions of primary CVRs}



For each CVR, the probability of a collision with another CVR per unit time is $n\sigma_1 \delv $, where  $\delv = v\delR/R$ (as $v \propto R^{-1}$), and the cross-section is $\sigma = \sigma_1 ' R^2$, where $\sigma_1 ' \sim 1$. Naive geometric considerations suggest $\sigma_1 ' = 4\pi $, although further research into this problem is necessary \cite{Paul_sigma}. 
If each collision effectively removes two primary CVRs from the coherent beam, the total number of removed CVRs per unit time and unit volume:
\begin{equation}f_a = n^2 \sigma_1 \delv = \frac{\Lambda-1/2}{4\pi}\kappa \sigma_1 ' \delR n^2.\label{f1}\end{equation}
After using Eqs.\,\ref{nablav}\&\ref{tau1}, and $f=f_a$ from the above expression, Eq.\,\ref{RateEquation} becomes:
\begin{equation}\frac{\dif n}{\dif t}  -  \frac{n}{2t} + \frac{d \sigma_1 ' \delR R_1}{2\tau_1} n^2 = 0.\label{RateEquation1}\end{equation}
Its asymptotic solution is 
\begin{equation}n = \frac{3}{\sigma_1 ' \delR d}R_1^{-1}\frac{\tau_1}{t}.\label{Solution1}\end{equation}  
At $t=\tau_1$, 
\begin{equation}
n_1R_1 = \frac{3}{\sigma_1 ' \delR d}. 
\label{nR1}
\end{equation}
The RHS is independent of both the injection intensity and $U$. For $\sigma ' =4\pi$  and $\delR = 0.5$\,$\mu$m, it is equal to $4\times10^3$\,cm$^{-2}$, which is a factor of 8 greater than the experimental value of $5\times10^2$\,cm$^{-2}$. There might be several reasons for such a discrepancy. Firstly, the cross-section for the effective removal of CVRs from the coherent beam of similar orientation and velocities might be several times greater than the geometric guess $\sigma = 4\pi R^2$. Secondly, in this model we restricted all CVRs to motion along the $x$-axis with only a spread of radii; in reality they also have a small random spread of directions of motion caused by the conserved transverse component of the impulse of the initial CVRs -- our estimates show that its contribution to the frequency of collisions is comparable with that calculated here.     

\subsection{Multiple pile-ups of primary CVRs}

Another process that should further limit the density of primary CVRs is the removal of all CVRs that bump into the secondary vortex rings formed upon any binary collision discussed above. 
Large secondary CVRs appear with each primary collisions at a rate, per unit volume and time, 
\begin{equation}\frac{f_a}{2} = \frac{(\Lambda-1/2)\kappa \sigma_1 ' \delR}{8\pi} n^2. \label{f12}\end{equation}
As these larger rings are considerably slower than the primary ones, they block all the primary CVRs from behind within the effective cross-section $\sigma_2 = \sigma_2 ' R^2$, where $\sigma_2 ' \sim 1$, and the upper limit on its geometrical value is $\sigma_2 ' = 4\sqrt{2}\pi  \approx 18$ \cite{sigma2}. The typical number of primary CVRs lost in such a capture (the ``multiplication factor'')  is $N_2 \sim n\sigma_2\frac{\Delta x}{2}$. The rate of losses per unit volume and time is hence
\begin{equation}f_b = N_2\frac{f_a}{2} = \frac{(\Lambda-1/2) \kappa \sigma_1 ' \sigma_2 ' \delR R_1^2 \Delta t d}{32\pi \tau_1^{3/2}}  t^{1/2} n^3.\label{f2}\end{equation}
With $f=f_b$, Eq.\,\ref{RateEquation} becomes 
\begin{equation} 
\frac{\dif n}{\dif t}  -  \frac{n}{2t} + \frac{(\Lambda-1/2) \kappa \sigma_1 ' \sigma_2 ' \delR R_1^2 \Delta t d}{32\pi \tau_1^{3/2}} t^{1/2} n^3 =0.
\label{RateEquation2}\end{equation}
Its asymptotic solution is:
\begin{equation} n =  \left( \frac{40\pi }{(\Lambda-1/2) \kappa \sigma_1 ' \sigma_2 ' \delR  \Delta t d} \right)^{1/2} R_1^{-1} \left(\frac{\tau_1}{t}\right)^{3/4}.\label{Solution2}\end{equation}
For $t=\tau_1$ this gives
\begin{equation}
n_1R_1=\left( \frac{40\pi}{(\Lambda-1/2)\kappa \sigma_1 ' \sigma_2 ' \delR \Delta t d} \right)^{1/2},
\label{nR2}
\end{equation}
independent of the injection intensity and $U$. For $\sigma_1 ' = 4\pi$, $\sigma_2 ' = 4\sqrt{2}\pi$, $\delR = 0.5$\,$\mu$m, $\Delta t = 0.2$\,s and $d=4.5$\,cm, this produces $n_1R_1 = 1.0\times 10^3$\,cm$^{-2}$, which  is only factor of 2 greater than the experimental value. Furthermore, as we only considered the effect of longitudinal variations of the velocities $\delv$ of primary CVRs but disregarded the comparable effect from their transverse velocities, the agreement is quite good.


\subsection{Conclusion}
To conclude, the solutions, $n_1R_1=const$, found for either of the simple models of removal of primary vortex rings upon collisions, Eq.\,\ref{nR1} and Eq.\,\ref{nR2}, qualitatively agree with our experimental observations. Moreover, the latter is actually in a reasonable quantitative agreement.  
Thus multiple pile-ups explain the observed universal value of the product $n_1R_1$ when CVRs collide frequently due to their large numbers and large radii. 
Further experiments with varyable flight path $d$, duration of injection $\Delta t$ and distribution of CVRs' radii $\delR$, as well as numerical simulations of the interaction between CVRs, will help improve our understanding of the microscopic processes that occur within beams of unidirectional vortex rings. 




\begin{thebibliography}{99}
\bibitem{1-d} In our experiments, CVRs' motion very near the injector is not quite one-dimensional, but the number of reconnections is small there, hence the end result should not be strongly affected by this approximation.
\bibitem{Paul_sigma} Our preliminary calculations of the interaction between two parallel vortex rings of slightly different radii
show that reconnections occur with the cross-section $\sigma _1 \approx 0.6 R^2$, while a noticible perturbation of the directions of motion and velocities (even without reconnections) occurs with the cross-section even larger than $\sigma _1 = 4\pi R^2$.
\bibitem{sigma2} We assume that, upon a collision of two primary CVRs of radii $R$ each carrying impulse $\rho\kappa\pi R^2$, they merge into a circular loop with impulse $2\rho\kappa\pi R^2$ and hence radius $\sqrt{2}R$.
\end{thebibliography}
\end{document}